# Compact Muon Production and Collection Scheme for High-Energy Physics Experiments

Diktys Stratakis

*Brookhaven National Laboratory, Upton NY 11973, USA*

David V. Neuffer

*Fermi National Accelerator Laboratory, Batavia, IL 60510, USA*

(Dated: September 09, 2014)

## Abstract

The relative immunity of muons to synchrotron radiation suggests that they might be used in place of electrons as probes in fundamental high-energy physics experiments. Muons are commonly produced indirectly through pion decay by interaction of a charged particle beam with a target. However, the large angle and energy dispersion of the initial beams as well as the short muon lifetime limits many potential applications. Here, we describe a fast method for manipulating the longitudinal and transverse phase-space of a divergent pion-muon beam to enable efficient capture and downstream transport with minimum losses. We also discuss the design of a handling system for the removal of unwanted secondary particles from the target region and thus reduce activation of the machine. The compact muon source we describe can be used for fundamental physics research in neutrino experiments.

## I. INTRODUCTION

The relative immunity of muons to synchrotron radiation [1] due to their large rest mass ($m_\mu$=105.7 MeV/c$^2$) suggests that they may be used in place of electrons for fundamental high-energy physics research as well as for various industrial and medical applications. For instance, accelerated muon beams can enable unique element analysis via muonic X-rays and muon radiography [2]. In addition, compact muon accelerators are desired for medical [3] and material detection applications [4]. Moreover, muon accelerators are being explored for a Neutrino Factory [5] and a Muon Collider [6]. However, the short lifetime of muons – 2.2 µs in the rest frame – makes the transport of a muon beam very challenging technologically [7].

Muons are charged particles with mass between the electron and proton and can be produced indirectly through pion decay by interaction of a high-energy (multi-GeV), high-power (1-4 MW) proton beam with a stationary target [8]. The muon yield is fractionally small, with large angle and energy dispersion, so that efficient collection is necessary in all dimensions of phase-space [9-11].

In this paper the production of pions, their decay into muons and the survival of muons during transport are studied. We present a method [12] for manipulating the longitudinal and transverse phase-space in order to efficiently capture and transport a muon beam from the production target towards the accelerator chain. In that method, a set of properly tuned rf cavities captures the muon beams into strings of bunches and aligns them to nearly equal central energies, and a following set of rf cavities with absorbers cools them by a factor of three in transverse emittance. With the aid of numerical simulations, we show that our proposed muon collection scheme can simultaneously capture species of both signs with a notable rate of 0.12 muons of each sign per incident 8 GeV proton. We systematically analyze the sensitivity in performance of the channel against key parameters such as the number of cavities, accelerating gradient and magnetic field. The compact muon source we describe can be used for fundamental physics research in neutrino experiments [5, 6].

It is important to emphasize that the aforementioned method of pion production creates a significant background of electrons and protons which may result in heat deposition on superconducting materials and activation of the machine, preventing manual handling. Therefore, in this paper we discuss the design of a secondary particle handling system [13]. The system comprises a solenoidal chicane that filters high momentum particles, followed by a proton absorber that reduces the energy of all particles, resulting in the rejection of low energy species that pass through the chicane. Our findings indicate that such a system can successfully remove high-energy particles (>1 GeV/c) and thus significantly reduce activation of accelerator components. We detail the design and optimization of the system and its integration with the rest of the muon capture scheme.

## II. MUON CAPTURE SCHEME

The main components of the muon capture scheme are illustrated in Fig. 1. In the subsections below we review those components:

### A. Target and Decay Channel

In the present baseline configuration, the input beam was taken to be composed of a monochromatic 8 GeV proton beam with a 2 ns time spread incident on a liquid mercury target in the bore of a 20 T solenoid [14]. The proton beam and mercury jet are tilted with respect to the solenoidal field magnetic axis, so that non-interacting particles impinge on the mercury jet collection pool which acts as the proton dump [15]. The interaction of the beam with the target was modeled using MARS [16], assuming $10^6$ incident protons on the target. We then propagate the produced particles using ICOOL [17], a code that can track all relevant physical processes (e.g. energy loss, straggling, multiple scattering) and includes muon decay. After the target, the field tapers from 20 T to 2.0 T, collecting both positive and negative particle species (see Fig. 1). Then, the field continues at 2 T for another 15.25 m. Note that the aperture radius in the decay channel as well as in all subsequent sections is assumed to be 0.25 m. Particles that hit the wall are considered lost.

A major improvement in our present design is that the magnetic field drops from its peak value down to 2 T over 6.00 m rather than the 14.75 m considered in previous studies [6]. In ICOOL, the field is modeled based on an inverse cubic profile [18], from which the off-axis fields were approximated using series expansions. The axial solenoid field profile is illustrated in Fig. 2(a), with z=0 being the location of the target. Quantitatively, the effect of the solenoidal field profile on the lattice capture efficiency can be evaluated by calculating the muon yield, which is defined as the number of particles that fall within a reference acceptance, and this yield approximates the expected acceptance of the downstream accelerator. For the Neutrino Factory case, which is the one assumed in this study, the transverse normalized acceptance is 30 mm and the normalized longitudinal acceptance is 150 mm [19]. Figure 2(b) shows the relative muon yield for the two axial profiles under consideration. Clearly, the new profile (red curve) increases the final muon yield by ~8% and this is a direct consequence of the fact that a short taper delivers a denser distribution of muons in longitudinal phase-space, which permits a more effective bunch formation in the buncher and phase-rotator sections further downstream [20].

*B. Chicane and Absorber*

In addition to the desirable pions which will eventually decay into muons, a large flux of protons of all energies comes off the target up to the energy of incoming protons. Additionally, relatively low-energy electrons and pions are produced. Without collimation, this flux is lost on the capture channel apertures at kW/ m levels, much larger than the approximately 1 W/m desired to ensure "hands on" maintenance [21]. Rogers [13] proposed a solenoid chicane to eliminate the high energy protons and other unwanted secondaries and an absorber to deal with the remaining low energy particles. The chicane is a bent solenoid system. High momentum particles are not strongly deflected by the bend solenoid and are lost in or near the chicane and collimated on shielded walls. Lower momentum particles are strongly focused by the solenoid and follow the chicane with little orbit distortion.

For our system the solenoidal chicane is placed at the end of the 15.25 m decay drift. The chicane bends out by $\theta$ =15.0 deg. over $L$ =6.5 m and back by the same angle over 6.5 m more [Fig. 3(a)]. The chicane magnetic field is simulated in ICOOL using a toroidal model. Specifically, the magnetic field in each chicane section is modeled with a purely longitudinal field of $B_0(1 + \theta x/L)$, where $x$ is the horizontal coordinate, positive away from the center of curvature and $B_0 = 2$ T. There is a constant 2 T solenoidal field downstream from that point. Note that simulations using a field map generated by a set of coils produced the same results [22].

The chicane can process particles of both signs and its action on the incident protons is displayed in Fig. 3(b). A glance in Fig. 3(b) indicates that high momentum particles (> 1 GeV/c) are lost in or near the chicane and collimated on shielded walls. The remaining particles pass through a 10 cm Beryllium absorber which removes almost all of the remaining low energy protons. The absorber was found to stop pions before they decay to muons and was therefore moved 30.1 m downstream of the chicane. An additional 21.15 m was added after the absorber to extend the beam distribution and obtain the energy position correlation needed for bunching and phase-rotation. The chicane/absorber system was found to reduce the downstream energy deposition by more than an order of magnitude over a muon collection scheme without it [22].

Figure 4 shows the evolution of the muon beam longitudinal phase-space starting from the target [Fig. 4(a)]. As the beam drifts away, pions decay into muons, and the bunch lengthens, developing a "high energy head" and a "low energy tail". At the same time, the chicane chops away muons with momentum > ~1 GeV/c [Fig. 4(b)]. The separation of the particles that develops is given by:

$$\delta(ct_i) = L\left(\frac{1}{\beta_i} - \frac{1}{\beta_0}\right), \tag{1}$$

where $\delta ct$ indicates the time delay from a reference particle of speed $\beta_0$, L is the distance from the production target and $\beta_i$ is the particle longitudinal speed.

C. Buncher and Phase-Rotator

The decay channel is followed by a 21 m buncher section in which the gradient of the rf system gradually increases and the beam is captured into a string of bunches with different energies. The rf frequency decreases along the length of the buncher, with the constraint that the phase difference between two reference particle momenta, $P_0$ and $P_N$, remains a fixed number $N_B$ of wavelengths as the beam propagates through it, i.e.,

$$N_B \lambda_{rf}(L) = L \left( \frac{1}{\beta_N} - \frac{1}{\beta_0} \right). \tag{2}$$

where $\beta_0$ and $\beta_N$ are the velocities of the reference particles at momentum $P_0$ and $P_N$, respectively. Following this procedure, the reference particles and all intermediate bunch centers remain at zero crossings of the rf wave throughout the buncher. For the present design, $N_B$ is chosen to be 12, $P_0 = 250$ MeV/c and $P_N = 154$ MeV/c with the intent of capturing particles within 50 to 400 MeV energy range. With these parameters, the rf frequency at the beginning of the buncher section is 490 MHz and at the end falls to 365 MHz. In the bunching system, 56 normal conducting pillbox-shaped rf cavities are employed, each having a different rf frequency and the rf gradient G increases linearly by

$$G = G_0 \left( \frac{L}{L_B} \right), \tag{3}$$

where $L_B$ is the buncher length. In the above expression, $G_0$ is the gradient at the end of the buncher and equals 14.30 MV/m. The gradual increase of voltage enables a somewhat adiabatic capture of muons into separated bunches. The buncher consists of 28 cells, 0.75 m each, containing two 0.25 m long cavities. To keep the muon beam focused, a constant 2.0 T field is maintained through the section. A schematic layout of 3 buncher cells is shown in Fig. 5.

Once the beam leaves the buncher, it consists of a train of bunches with different energies [Fig. 4(c)]. The beam then is phase-rotated with a second string of cavities with decreasing frequencies, but with constant accelerating gradient. The frequencies are chosen so that the centers of the low- energy bunches increase in energy, while those of the high-energy bunches decrease. The algorithm [12] used for setting this condition is to keep the first reference particle at fixed momentum while uniformly accelerating the second reference particle through the rotator section, so that it attains the first particle's energy at the end of the channel. This is accomplished by increasing slightly the phase shift between the reference particles by $N_R = N_B + 0.045$. With this condition, the bunches are aligned into nearly equal energies over the 24 m length of the rotator. The rf gradient is kept fixed at 20 MV/m while the rf frequency drops from 364 to 326.5 MHz. At the end of the rotator the reference particles are at the same momentum ~245 MeV/c and the rf frequency is matched to 325 MHz. Similar to the buncher, the rotator cell is 0.75 m in length, contains two rf cavities 0.25 m long each and the rotator includes 32 cells with a total of 64 rf cavities. The 2.0 T focusing field continues throughout the rotator section. The longitudinal phase-space distribution of the beam at the exit of the phase-rotator is shown in Fig. 4(d). For muon accelerator applications [5, 6, 23], only 21 bunches (shown within the red box) are used for subsequent cooling and acceleration.

A critical feature of the muon production, collection, bunching and phase rotation system is that it produces bunches of both positive and negative muons at roughly equal intensities. This occurs because the focusing systems are solenoids which focus both signs, and the rf systems have stable acceleration for both signs, separated by a phase difference of π. To illustrate this point, in Fig. 6 we plot a few interleaved positive and negative muon bunches at the exit of the phase-rotator.

The presence of magnetic fields overlapping rf cavities has been identified as a technical risk that may reduce the capture efficiency of the muon selection system due to a reduction in the peak gradient that can be achieved in the rf cavities [24-26]. Therefore, in Fig. 7 we examine the consequences of using lower rf gradients in the phase-rotator section. As noted earlier, the lattice efficiency can be calculated by counting the number of simulated particles that fall within a reference acceptance, which approximates the expected acceptance of the downstream accelerator. As the plot indicates, a lower gradient would clearly hurt the final muon yield. For instance, a drop from the current baseline gradient of 20 MV/m to 15 MV/m would reduce the number of accepted muons by more than 10%.

The present design so far assumes a continuously decreasing frequency where each cavity is different resulting to a total of 120 rf frequencies in the buncher and phase-rotator combined. In a more realistic implementation, the cavities need to be grouped into a smaller number of frequencies. Thus, a critical question is how many rf cavities can be grouped into a certain frequency without loss in performance.

Table I displays the muon yield when the cavities are grouped into 1-pair (baseline), 4-pair and 8-pair frequencies. The simulations suggest that if the cavities are grouped into a pair of four, which corresponds to 30 discrete frequencies, the relative muon yield is reduced by ~9%. However, if the cavities are grouped into 8-pair configuration (15 discrete frequencies) the resulting muon yield will drop by more than 20%. From the above results, it appears preferable to combine the cavities into groups of four and Table II has the required frequencies and rf cavity gradients to achieve this goal.

D. *Matching section*

While the magnetic field is constant at 2.0 T in the decay channel, buncher and phase-rotator section, the field in the subsequent cooling channel is generated by alternating solenoid coils of ±3 T strength. Without proper matching between the region of constant and alternating solenoid fields, emittance growth and severe particle loss can occur. We consider a sequence of 9 solenoid coils to carry out the match [27]. The magnet settings were optimized using a standard Nelder-Mead algorithm [28] with the objective to maximize the muon yield in the subsequent cooling channel. In our analysis we consider that the matching section is followed by a 4-dimensional cooling channel with block absorbers which is described in more detail in the next section.

Figure 8 shows the number of accepted muons (muon yield) along different parts of the channel before and after the match. It becomes evident that the inclusion of a matching section adds a 8% to the total gain. Acceptance is maximal at ~ 0.120 muons per initial 8 GeV proton at z=240 m (~100 m of cooling).

E. *Cooling Section*

Upon exiting the matching section, the muons enter a ~ 100 m long ionization cooling channel [29] consisting of rf cavities, disc absorbers, and alternating solenoids for focusing. The channel consists of a sequence of ~70 identical 1.5 m long cells. Each cell contains 2 doublets of 0.25 m long 325 MHz pillbox rf cavities with 0.25 m spacing between the doublets and a 1.5 cm thick Lithium Hydride (LiH) discs at the ends of each doublet (four per cell). A schematic layout of one cooling cell is shown in Fig. 9. The LiH provides the energy loss material for ionization cooling. Each cell contains two solenoid coils with opposite polarity, yielding an approximately sinusoidal variation of the magnetic field with a peak value of ~ 3 T on axis. The axial length of the coil is 15 cm, with an inner radius of 35 cm, an outer radius of 50 cm and a current density of 105.6 A/mm$^2$. In ICOOL, we generated 2D cylindrical field maps by superimposing the fields from all solenoids in the cell and its neighbor cells. Based on the simulation results, the cooling channel reduces the transverse phase-space volume by a factor of 3 with a final rms normalized emittance equal to 6.0 mm.

## III. SUMMARY

Beams of accelerated muons are of great interest for fundamental high-energy physics research as well as for various industrial and medical applications. Muons are produced indirectly through pion decay by interaction of a charged particle beam with a target. However, the muon yield is fractionally small, with large angle and energy dispersion, so that efficient collection in all dimensions in phase-space is necessary. Here, we have described a fast method for manipulating the longitudinal and transverse phase-space with the purpose to efficiently capture and transport a muon beam from the production target towards the accelerator chain. In that method, a set of properly tuned rf cavities captures the muon beams into strings of bunches and aligns them to nearly equal central energies, and a following set of rf cavities with absorbers cools them by a factor of three in transverse emittance. With the aid of numerical simulations we found that our present muon collection scheme is capable of capturing simultaneously muons of both signs with a notable rate of 0.12 muons per initial 8 GeV incident proton. In addition, we discussed a conceptual design of a handling system for the removal of unwanted secondary particles from the target region (such as electrons and protons). We showed that such a system can successfully remove particles with momentum > 1 GeV/c and thus significantly reduce activation of the machine. The compact muon source described here can be used for fundamental physics research in neutrino experiments [5, 6], muon radiography [2] to study industrial machinery and medical research such as functional brain studies through muon-spin relaxation technique [3].

## ACKNOWLEDGEMENTS

The authors are grateful to J. S. Berg, H. Kirk, H. K. Sayed, and R. B. Palmer for useful discussions. This work is supported by the U.S. Department of Energy, Contract no. DE-AC02-98CH10886

15

# TABLES AND FIGURE CAPTIONS

**Table I: The effect of cavity grouping to discrete frequencies on the relative muon yield: one cavity per frequency (first row), four cavities per frequency (second row) and eight cavities per frequency (third row). In our analysis, the maximum muon yield of the 1-rf pair case is normalized to one.**

| Lattice | No. of Buncher frequencies | No. of Rotator frequencies | Total No. | Relative Muon yield |
|---|---|---|---|---|
| 1-rf pair | 56 | 64 | 120 | 1 |
| 4-rf pair | 14 | 16 | 30 | 0.91 |
| 8-rf pair | 7 | 8 | 15 | 0.76 |

**Table II: Buncher and phase-rotator rf parameters.**

| Buncher rf frequency (MHz) | Buncher rf gradient (MV/m) | Rotator rf frequency (MHz) | Rotator rf gradient (MV/m) |
|---|---|---|---|
| 493.71 | 0.30 | 363.86 | 20 |
| 482.21 | 1.24 | 357.57 | 20 |
| 470.27 | 1.95 | 352.20 | 20 |
| 458.40 | 3.38 | 347.59 | 20 |
| 448.07 | 4.45 | 343.65 | 20 |
| 437.73 | 5.52 | 340.27 | 20 |
| 427.86 | 6.60 | 337.39 | 20 |
| 418.43 | 7.67 | 334.95 | 20 |
| 409.41 | 8.74 | 332.88 | 20 |
| 400.76 | 9.81 | 331.16 | 20 |
| 392.48 | 10.88 | 329.75 | 20 |
| 384.53 | 11.95 | 328.62 | 20 |
| 376.89 | 13.02 | 327.73 | 20 |
| 369.55 | 14.30 | 327.08 | 20 |
|  |  | 326.65 | 20 |
|  |  | 326.41 | 20 |

**Fig. 1:** Schematic representation of a muon collection system for high-energy physics experiments, showing a target, a drift, a chicane, an rf buncher, a phase-energy rotator, and a cooler section. Pions are produced by protons on a target at the beginning of the drift, decay to muons in the drift, and while lengthening in time, the buncher and rotator change the muons into a string of bunches with nearly equal energies.

**Fig. 2:** (a) Axial magnetic field profile within the muon capture region that is used in the present configuration (red curve) vs to the profile used in older studies (blue curve) [9] ; (b) relative muon yield per incident proton for each profile. The target is placed at z=0. The present configuration uses a shorter taper profile which results to an 8% increase on the total accepted muons.

**Fig. 3:** (a) Schematic representation of the capture region, including a chicane to suppress energetic protons and an absorber for soft protons; (b) distribution of protons 34.25 m downstream of the target with (red) and without (blue) a chicane. The chicane removes high energy protons from the beam and thus significantly reduces activation of the machine.

**Fig. 4:** Evolution of the longitudinal phase-space along the drift, buncher, and phase-rotator within our proposed muon collection channel. (a) Pions and muons as produced at the target; (b) muons after the drift; (c) at the end of the buncher; and (d) at the end of the rotator. The present muon capture scheme produces 21 muon bunches (see red box) which are aligned into nearly equal energies that match the desired criteria for neutrino experiments [5, 6].

**Fig. 5:** Schematic layout of three lattice cells at the beginning of the buncher section.

**Fig. 6:** Distribution of particles in longitudinal phase-space at the phase-rotation end. $\mu^+$ are shown in black and $\mu^-$ in red. A critical feature of the present muon capture scheme is that it produces bunches of both signs at equal intensities.

**Fig. 7:** Dependence of the relative number of accepted muons per incident proton upon the average axial accelerating gradient of the phase-rotator rf cavities. Note that the muon yield when the rf gradient is 20 MV/m is normalized to 1.

**Fig. 8:** Muon yield along the muon capture channel before and after optimizing a matching section between the phase-rotator and cooler.

**Fig. 9:** Schematic layout of one lattice cell of the cooling section. The cooling section involves 70 identical cells and cools the transverse emittance by a factor of 3.

**Figure 1**

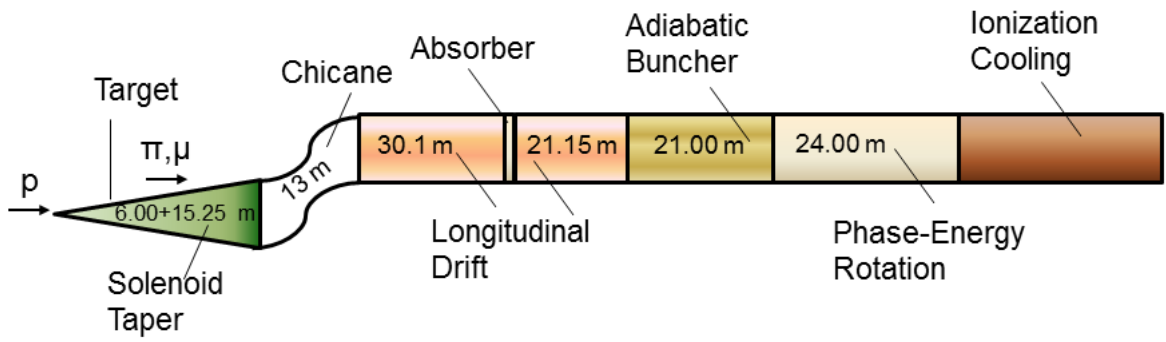

**Figure 2**

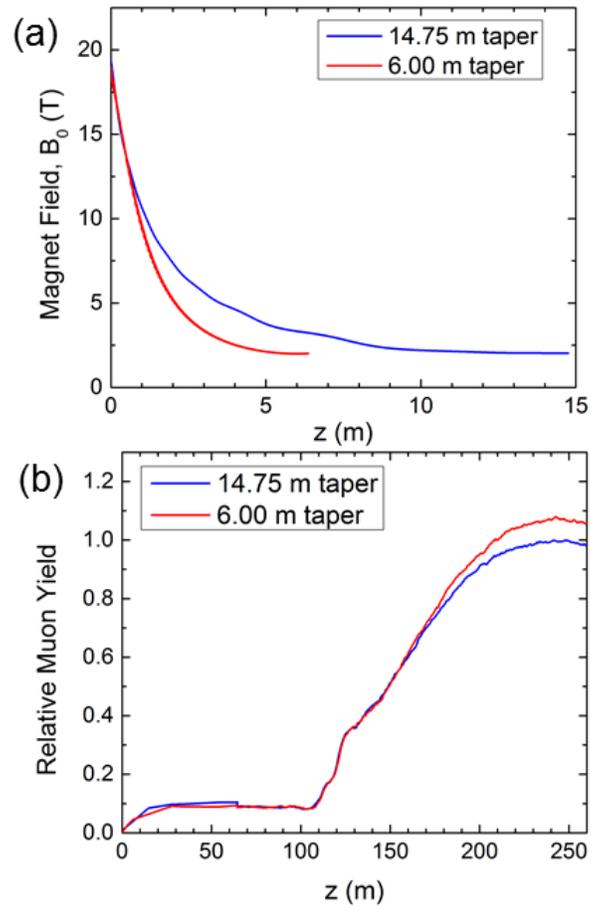

**Figure 3**

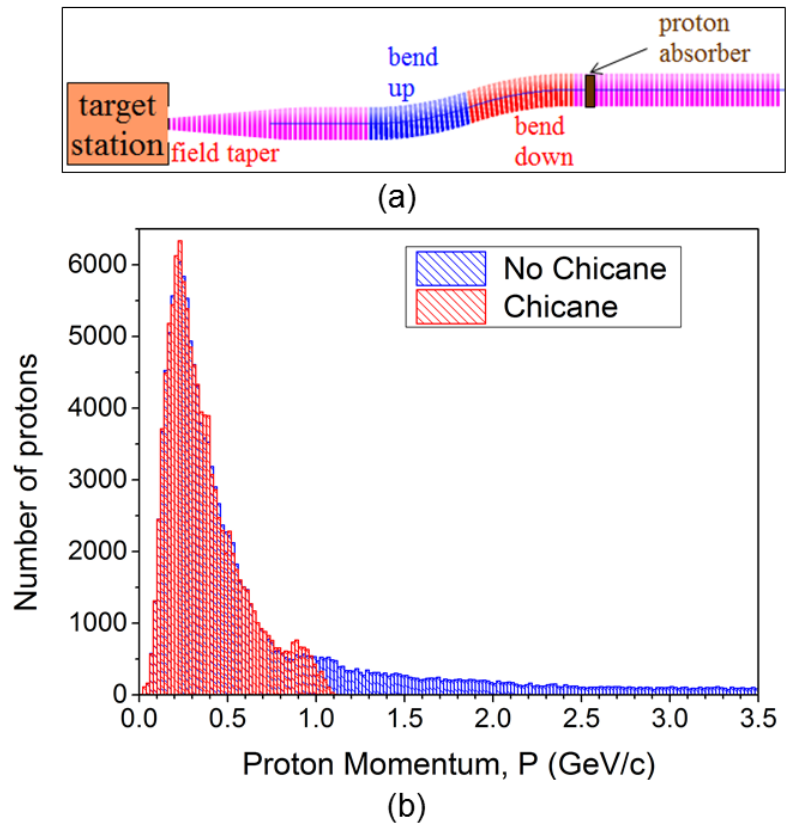

**Figure 4**

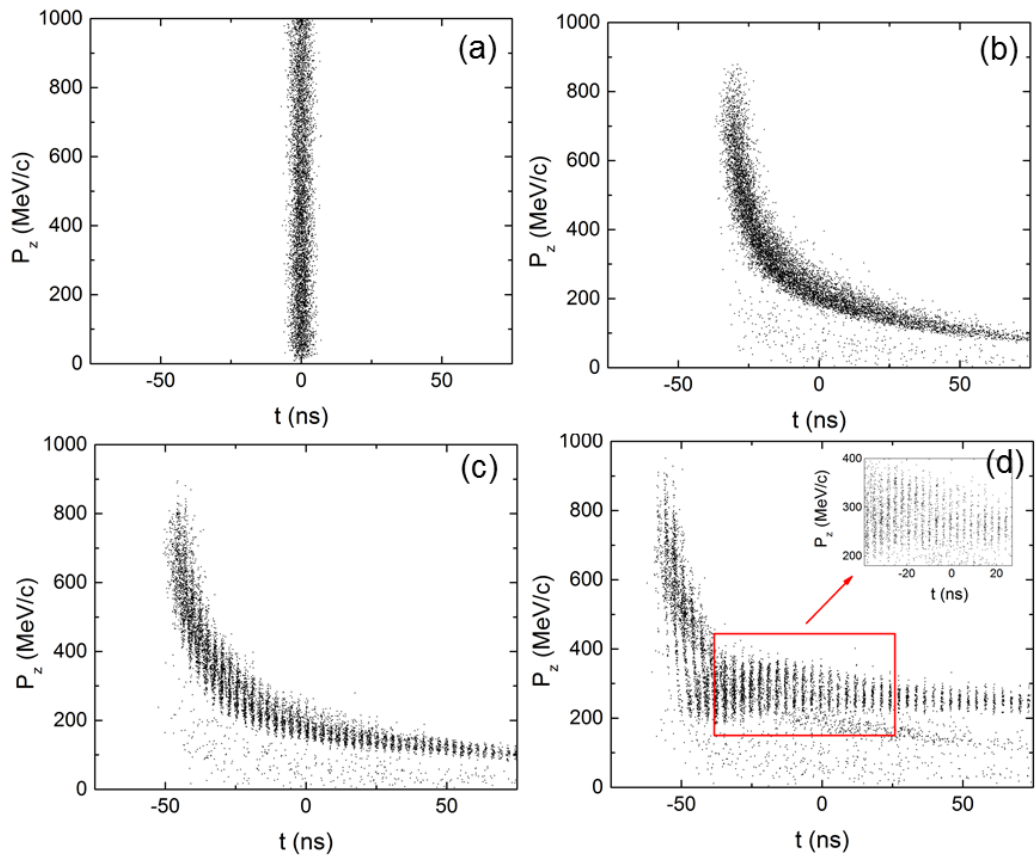

**Figure 5**

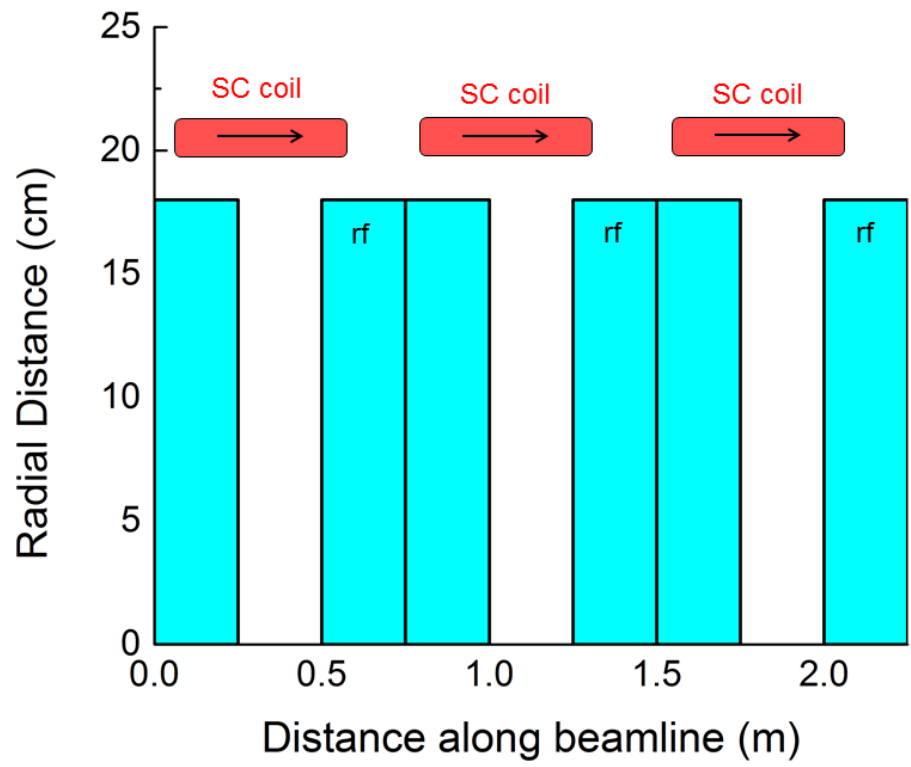

**Figure 6**

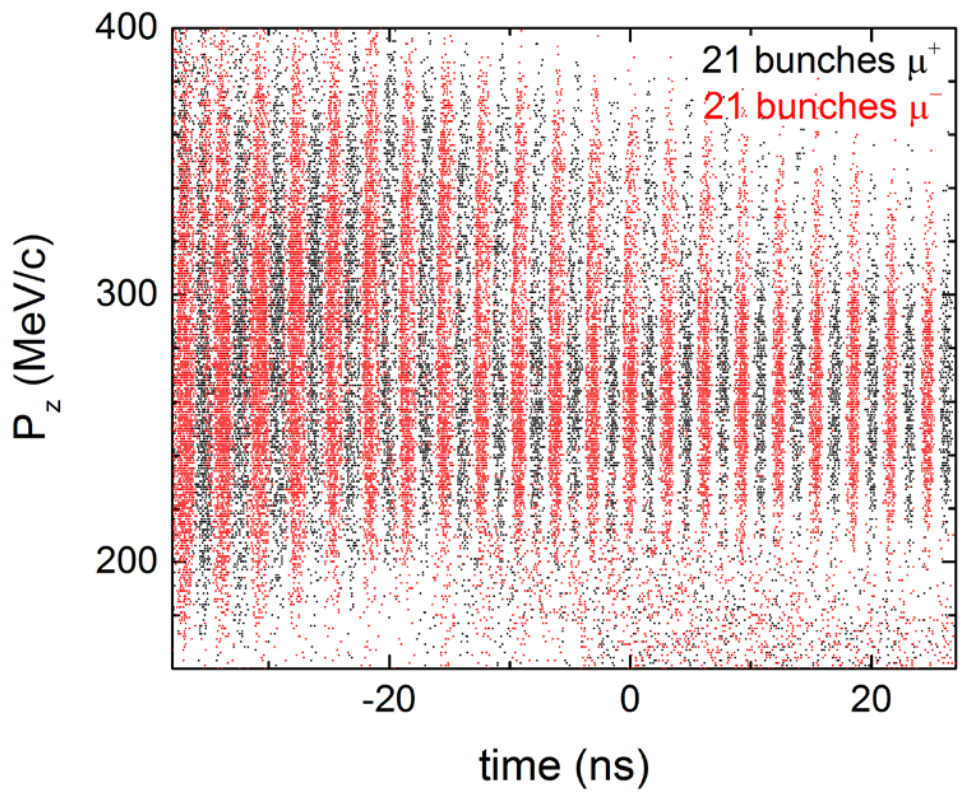

**Figure 7**

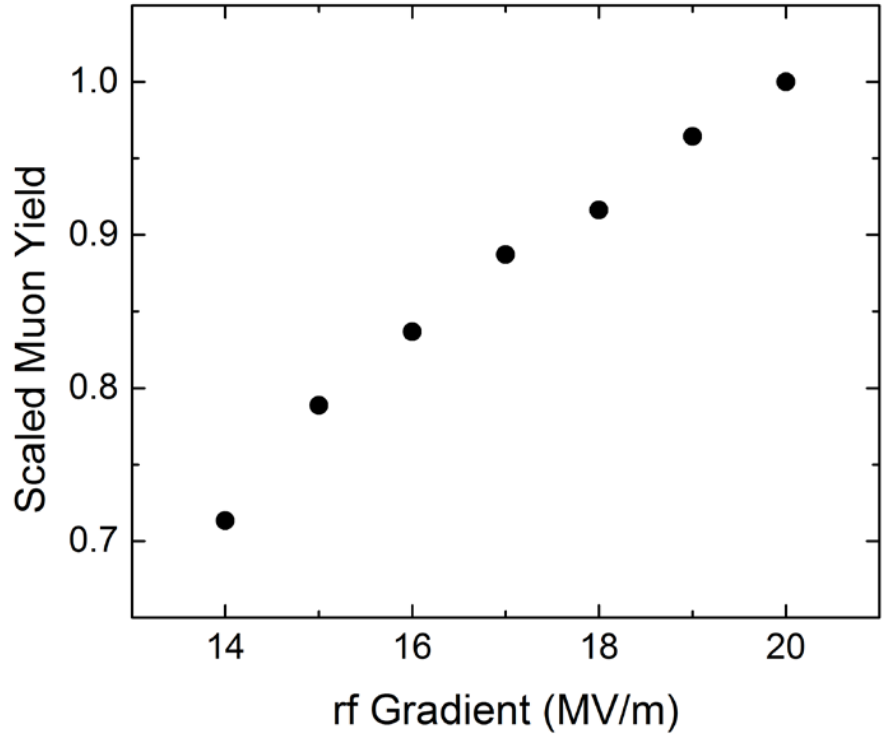

**Figure 8**

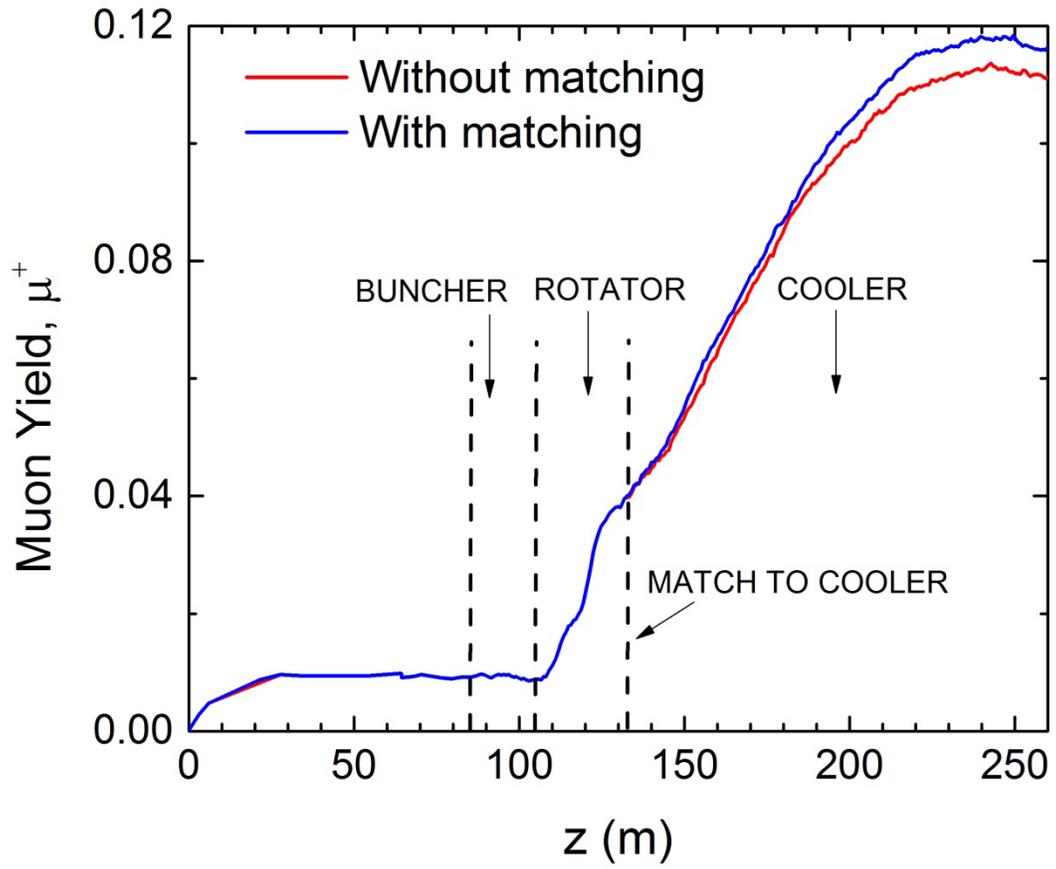

**Figure 9**

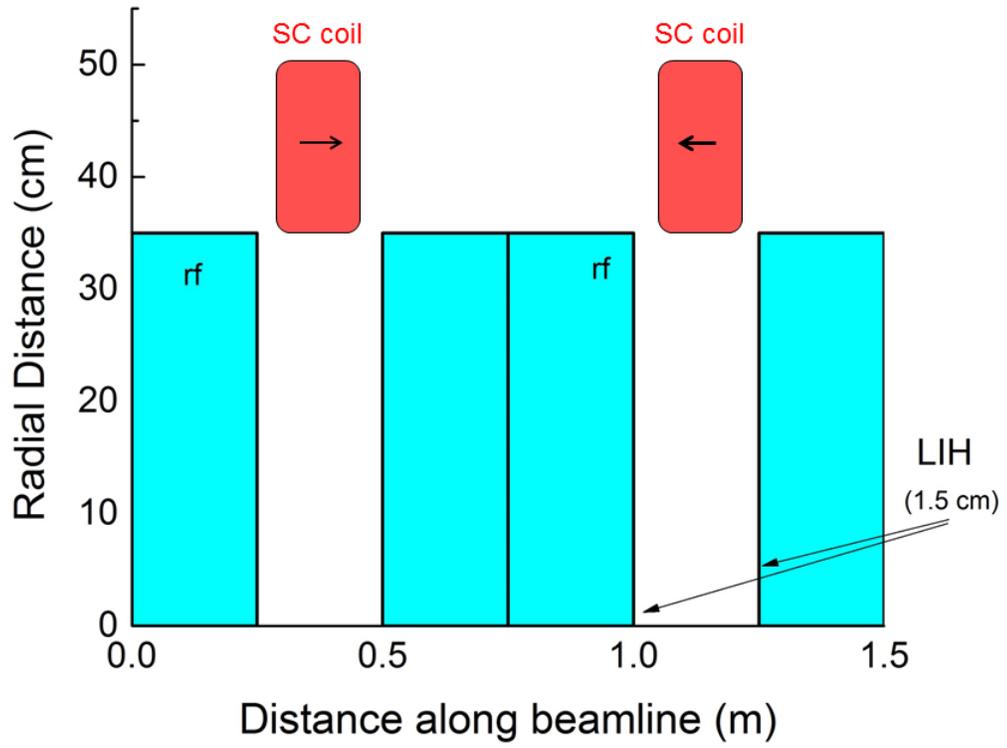